\def\eqnum#1{\eqno (#1)}

\documentclass[12pt]{article} \parskip 2pt

\usepackage{latexsym}

\begin{document}

\title{\bf
Classical approaches to Higgs mechanism
\vspace{10pt}}
\author{{\bf Assen Kyuldjiev\thanks{E-mail:
{\scriptsize KYULJIEV@INRNE.BAS.BG}. Supported by the Bulgarian
National Foundation for Science under grant $\Phi$-610.}
}\\[9pt]
{\normalsize Institute of Nuclear Research and Nuclear Energy,}\\ {\normalsize Tzarigradsko
chauss\'ee 72, Sofia 1784, Bulgaria}
\vspace{5pt}}

\maketitle

\begin{abstract}

The standard approach to Higgs mechanism is based on the existence of 
unitary gauge but, unfortunately, it does not come from a coordinate change 
in the configuration space of the initial model and actually defines a new 
dynamical system. So, it is a questionable approach to the problem but it is
shown here that the final result could still make sense as a Marsden-Weinstein 
reduced system. (This reduction can be seen as completely analogous to the 
procedure of obtaining the ``centrifugal" potential in the classical Kepler 
problem.)

It is shown that in the standard linearization approximation of the Coulomb 
gauged Higgs model geometrical constraint theory offers an explanation of the 
Higgs mechanism because solving of the Gauss law constraint leads to different 
physical submanifolds which are not preserved by the action of the (broken) 
global U(1) group.

\end{abstract}  \newpage

Despite the phenomenal success of the Standard Model, Higgs mechanism remains yet
experimentally unverified.  
The current presentation of the spontaneous symmetry breaking (SSB) is still not quite convincing.
It boils down to applying change of variables (which a closer inspection reveals not to be a
coordinate change), in order to rearrange the quadratic terms in the Lagrangian in a form
suggesting presence of certain particles and absence of others.
Elimination of dynamics along the action of the global symmetry group to be broken is done by
hand and without justification.
It would be interesting to reanalyse the problem from purely classical viewpoint without appealing to
the quantum mystique.

To be concrete we shall concentrate on the Lagrangian analysed in \cite{Higgs} in its form
highlighting ``radial-angular" decomposition:
$$L = -\frac14 F_{\mu\nu}F^{\mu\nu} -\frac12 \partial_{\mu}R\, \partial ^{\mu} \!R -\frac12
e^2 R^2 A_{\mu}A^{\mu} +e R^2 A_{\mu}\,\partial ^{\mu}\theta -\frac12 R^2
\partial_{\mu}\theta\, \partial^{\mu}\theta - V(R^2)$$
and the corresponding Hamiltonian (when we assume the Coulomb gauge condition
$\mbox{\boldmath $\nabla .\, A$} = 0$) is:
\begin{eqnarray}
H&=&\frac12 \left[ \pi_R^2 + (\nabla R)^2  + \pi_{\mathbf A}^2 + (\mbox{\boldmath $\nabla$}
\times \mathbf A)^2  - 2\, (\mbox{\boldmath $\nabla$} A_0)^2 + R^2 (e\mathbf A -
\mbox{\boldmath $\nabla$} \theta)^2 \right] + \nonumber \\ & & \mbox{}
+\frac{\pi_\theta^2}{2R^2} + eA_0 \pi_\theta + V(R^2)
\end{eqnarray}
where $\pi_R = \partial _0R$, $\pi_{\mathbf A} = \mathbf E$, $E_k = F_{0 k}$ and $\pi_\theta =
R^2 (\partial_0\theta - eA_0)$ and $(- + +\ \!+)$ metric signature is assumed.

Most of the treatments of Higgs model make use of the so called unitary gauge defining
$eB_{\mu}= eA_{\mu} - \partial_{\mu}\theta$ and rewriting the Lagrangian as:
$$L^\prime = -\frac14 F_{\mu\nu}F^{\mu\nu} -\frac12 \partial_{\mu}R\, \partial^{\mu} \!R
-\frac12 e^2 R^2 B_{\mu}B^{\mu} - V(R^2).$$

Like any Lagrangian, $L$ is a function on a tangent bundle $T{\cal M}$ and in this case the
configuration space ${\cal M}$ is the space of potentials $A_{\mu}$ and the fields $R$ and
$\theta$;  while $L^\prime$ is function on the tangent bundle of the new field  $B_{\mu}$ and the
field $R$. The mixing of variables on the configuration and tangent space means that $L^\prime$
presents a new dynamical system (possibly quite sensible one) which is not equivalent to the initial
one and which cannot be obtained by a mere coordinate change in the configuration space. The
standard explanation is that after a local gauge transformation the Lagrangian could be rewritten in
the new form but, in general, this is not an allowed procedure. A natural question arises whether
there is a more rigorous explanation of this recipe.

The present paper claims that the final result coincides with the Marsden-Weinstein (MW) reduction
\cite{MW} of the initial dynamical system (and is actually analogous to the treatment of the
classical Kepler problem leading to the ``centrifugal" potential). To remind, when we have a group
$G$ with a Lie algebra $\cal G$ acting on a symplectic manifold $P$ in a (strongly) Hamiltonian
manner and defining a Lie algebra homomorphism, we have a momentum mapping
${\cal J} : P \rightarrow \cal G$*$\,$
given by
$$\langle {\cal J}(p),a\rangle = f_a(p)\quad \forall a \in \cal G$$
where $f_a$ is the Hamiltonian function of the fundamental vector field defined by the action of $a$
and also satisfying $$[f_a,f_b] = f_{[a,b]} \quad \forall a,b \in \cal G$$
then the MW quotient manifold ${\cal J}^{-1}(\mu)/G_{\mu}$ has a unique symplectic structure
(provided $\mu$ is weakly regular value of $\cal J$ and the action of the isotropy group of $\mu$ --
$G_{\mu}$ on ${\cal J}^{-1}(\mu)$ is free and proper).
This is a powerful method for obtaining reduced dynamics on a symplectic space starting from
symplectic dynamical system with a symmetry. (We shall skip here any technicalities
like which group actions admit momentum maps,
$Ad$*-equivariance, clean intersections etc.)
In our case the group to be broken $U(1)$ acts as
$$\theta \rightarrow \theta + \phi \eqnum{2}$$
and the space ${\cal J}^{-1}(0)$
is the subspace defined by  $\pi_\theta \equiv  - R^2 B_0 = 0$. The group action quotiening of this
space amounts to elimination of any residual $\theta$-dependence.
Reducing the Hamiltonian (1) (and assuming that the Gauss law constraint $\Delta A_0 = e
\pi_\theta$ is solved) we obtain the Hamiltonian corresponding to $L^\prime$ in the $B_0 = 0$
gauge, in conformity with the standard interpretation.

It is noteworthy to analyse the more general case when we reduce by a nonzero value $\mu$ of the
dual algebra $\cal G$*$\,$. The MW quotiening would be
equivalent to fixing $\pi_\theta = const \neq 0$ and again factoring out
$\theta$ from  the phase space and the Hamiltonian. As a result the initial potential $V(x) =
-ax+bx^2$ with $a, b > 0$
will be modified by a $c x^{-2}$ term (with $c > 0$) and this will lead to higher values of the Higgs
mass without possibility for its vanishing.
(This could also add a new free parameter in possible future experimental testing of the Higgs
mechanism.)

This is actually not an explanation why $\theta$-symmetry is spontaneously broken---this is just a
more rigorous procedure for factoring out ($\theta$,$\,\pi_\theta$) degree of freedom and thus
eliminating the movement along $\theta$ which would be the dynamics typical for a massless field.
It is precisely this movement along the flat bottom of the potential surface which could lead to a
massless (Goldstone) field. One could still ask what prevents movement in this direction and hence
causing SSB. Being aware that SSB could only exist in systems with infinite degrees of freedom,
one may also wonder where this property is encoded in the above mentioned procedures.

A rigorous approach to these problems could be found e.g.\ in \cite{FS} where a structure of
Hilbert space sectors (HSS) is found in solutions of nonlinear classical relativistic field equations.
Each sector is invariant under time evolution, has a linear structure and is isomorphic to a Hilbert
space; and may be labeled by conserved dynamical charges. 
Different HSS define ``disjoint physical worlds" which could be considered as a set of
configurations which are accessible in a given laboratory starting from a given ground state
configuration. Then any group which maps a HSS into another HSS is spontaneously broken and
only ``stability" groups which map a HSS into itself would be proper symmetry groups.

Despite the nontriviality of existence of stable linear structures in the set of solutions of certain
nonlinear equations and the possibility to explain in principle the existence of SSB this approach
does not seem very practical.
Another possible route is offered by the use of geometrical constraint theory. Higgs model is a
beautiful example of a constrained system. Lagrangian does not depend on $\theta$ but only on
$\partial _{\mu}\theta$, thus allowing solutions with $\theta$-rotations but
($\theta$,$\,\pi_\theta$) degree of freedom remains coupled with the potential $A_{\mu}$. The
assumption that $\theta$-rotations are frozen (and consequent writing them off) obviously seems
ungrounded.

In what follows we shall return to our model in the Coulomb gauge.
This case was successfully tackled \cite{Bernst} by linearisation of the equations leading to
massive wave equation for $\theta$. Here we shall be interested not so much in the (linearised)
equations but in the symmetry breaking. We have primary constraint $\pi_0 \equiv \frac{\partial
L}{\partial(\partial_0 A_0)} = 0$ and the condition of its preserving gives the Gauss law constraint
equation $\Delta A_0 = e \pi_\theta$. To be precise, this equation is not the proper constraint --
the submanifolds determined by its solutions will give the surfaces on which the dynamics will be
constrained after factoring out the ($A_0$, $\pi_0$) degree of freedom.
Obviously, the solutions of the equation are
$$ A_{0} = G * e\pi_\theta  + f $$
where $G$  is a Green function for the Laplacian, $*$ denotes
convolution i.e. $(g*h)(x) = \int g(x\!-\!y) \:h(y) \:dy\;$ and
$f$ is any function satisfying $\Delta  f({\bf x},t) = 0$.
Solutions of this equation would define different physical submanifolds labeled by solutions of this
equation.
After differentiation we obtain
$$\dot{A_{0}} = eG * \dot{\pi_\theta}  + \dot{f} = -eG * \partial_k (R^2 B_k)  + \dot{f} $$
and taking into account that  $\partial_k (R^2 B_k) = \Delta (R^2 \theta)$ in the linearisation
approximation \cite{Bernst}, we have:
$$\dot{A_{0}} = eR^2\theta  + \dot{f}$$
This shows that we will have different dynamics on different physical submanifolds because the
general form of the ``massive wave equation'' for $\theta$ would be
$\Box \theta  = e^{2}R^2\theta  + e\dot{f}$           
(as long as linearisation approximation could be trustworthy).
More interestingly, transformations (2) does not act by lifting from the ``configuration space'':
$$
A_{0} \rightarrow  A_{0} \quad \dot{A_{0}} \rightarrow  \dot{A_{0}} + eR^2\phi
$$
and does not preserve the chosen submanifold. Transformation actions
of this kind are not very typical in  physics  (the standard  N\"other  theorem, 
for example, assumes  only  lifted transformations ). Thus we have a geometrical analog of
HSS and its origin could be traced to the requirements to the physical constrained
submanifolds in infinite dimensions \cite{AK}.(In this reference one could also see how this
phenomenon appears in an exactly solvable model.)


\section*{Acknowledgements}

The author is indebted to Prof. F. Strocchi for an illuminating and
inspiring discussion.

\end{document}